# Workplace Accidents and Self-Organized Criticality

John C. Mauro[1,*], Brett Diehl[1], Richard F. Marcellin, Jr.[2], and Daniel J. Vaughn[2]

[1]*Department of Materials Science and Engineering, The Pennsylvania State University,*

*University Park, Pennsylvania 16802, USA*

[2]*Science and Technology Division, Corning Incorporated, Corning, New York 14831, USA*

[*]*Corresponding Author:* jcm426@psu.edu

**Abstract:** The occurrence of workplace accidents is described within the context of self-organized criticality, a theory from statistical physics that governs a wide range of phenomena across physics, biology, geosciences, economics, and the social sciences. Workplace accident data from the U.S. Bureau of Labor Statistics reveal a power-law relationship between the number of accidents and their severity as measured by the number of days lost from work. This power-law scaling is indicative of workplace accidents being governed by self-organized criticality, suggesting that nearly all workplace accidents have a common underlying cause, independent of their severity. Such power-law scaling is found for all labor categories documented by the U.S. Bureau of Labor Statistics. Our results provide scientific support for the Heinrich accident triangle, with the practical implication that suppressing the rate of severe accidents requires changing the attitude toward workplace safety in general. By creating a culture that values safety, empowers individuals, and strives to continuously improve, accident rates can be suppressed across the full range of severities.

**Keywords:** Safety; Workplace Accidents; Accident Statistics; Power-Law Distribution; Self-Organized Criticality

**I. Introduction**

Despite significant progress over the past several decades, the occurrence of workplace accidents remains an unfortunate reality [1], with over 900,000 lost-time accidents reported in the U.S. in 2015 alone [2]. Workplace accidents span the full range of severity from minor injuries such as cuts or scrapes to fatalities. Of course, minor incidents occur much more frequently than major ones. In order to understand the origin of the most serious accidents, we must consider the statistics of rare events, an important subject in statistical physics with a wide range of applications across a variety of disciplines.

One of the most successful physical theories to describe the statistics of rare events is that of self-organized criticality. Self-organized criticality was proposed by the late Danish physicist Per Bak in 1987 as a means for explaining how complex behavior in nature can arise from relatively simple origins [3]-[5]. Bak and coworkers originally proposed the concept of self-organized criticality to explain the distribution of avalanche sizes in a sandpile [3]. Bak considered a sandpile where individual grains are slowly sprinkled on top. Most of the time, the individual grains just add to the growing pile. However, sometimes a grain creates a large enough perturbation to generate a landslide. The size of the landslide can be measured by the number of grains of sand involved, and smaller landslides occur much more frequently than larger landslides. Of course, the occurrence of larger landslides will have a greater impact on the structure of the sandpile. Given their much lower frequency of occurrence, the study of the probability of these larger landslides is an example of the statistics of rare events, which is addressed directly by Bak's theory of self-organized criticality.

The concept of self-organized criticality has several important implications. First, it entails the natural emergence of a power-law distribution to describe the frequency of events as a function of size,

$$N(x) = ax^{-k}, \qquad\qquad\qquad (1)$$

where $N$ is the number of occurrences of an event of size $x$, $a$ is a proportionality factor, and $k$ is a scaling exponent. The power-law distribution gives a characteristic straight line when plotted on log-log axes, i.e., when plotting the logarithm of the number of occurrences versus the logarithm of the avalanche size, as

$$\log N = \log a - k \log x. \qquad\qquad\qquad (2)$$

With such a log-log plot, the slope of the line reveals the value of the scaling exponent, $k$.

A second major insight from self-organized criticality is that all of the observed critical events have a common underlying origin, independent of their size. For the case of sandpiles, each landslide is caused by a perturbation induced by an individual falling grain of sand. The larger landslides occur much less frequently than the smaller landslides, but they all have the same fundamental cause. The term "self-organized" is used because the same scaling behavior emerges naturally, regardless of variation in the individual parameters of the experiment or the model.

Finally, a third major insight from self-organized criticality is its widespread applicability across a diverse range of fields, including physics, geology, materials science, biology, mathematics, economics, and even sociology [6]-[7]. An archetypal example from geology is the statistics of earthquake magnitudes [8]. All earthquakes have the same fundamental origin, i.e., a rock breaking along a fault line, which generates seismic waves. Low-magnitude earthquakes occur frequently, although we usually do not notice since they are too small to be felt. The earthquakes of greater concern are the more rarely occurring high-magnitude earthquakes that can create massive damage and loss of human life. The probability of these high-magnitude earthquakes occurring is another example of the statistics of rare events. Since all earthquakes have the same underlying origin, the frequency of earthquakes of varying magnitude follows a power-law distribution, the same distribution that describes the statistics of landslides in Bak's study of sandpiles. In this manner, both landslides and earthquakes are examples of self-organized criticality.

Other examples of phenomena described by self-organized criticality include forest fires, river flows, climatology, disease epidemics, financial markets, and social networks [6]-[7]. Self-organized criticality also describes the evolution of complex proteins in the field of biology [9]-[10]. Within materials science, self-organized criticality explains the existence of optimized stress-free glasses that exhibit a minimum in aging behavior [11]. In the social and economic sciences, self-organized criticality explains the probability distributions of wars and stock market crashes [6].

Here we propose that workplace accidents are another example of self-organized criticality. Workplace accident data from the U.S. Bureau of Labor Statistics reveal a power-law distribution of accident rates as a function of the number of days lost from work, a quantitative measure of the severity of an accident. As a characteristic feature of self-organized criticality, this power-law scaling points to a common underlying cause of workplace accidents, independent of their severity and the particular labor category in which they occur. We propose that the most effective means of reducing the rate of severe accidents in the workplace is to reduce the accident rate overall by creating a culture that incorporates safety as one of its core values. This result is consistent with the approach advocated by Heinrich [12] and provides a scientific basis for Heinrich's empirically derived accident triangle.

## II. Methods

The data are obtained from the U.S. Bureau of Labor Statistics (U.S. Department of Labor), gathered through the Survey of Occupational Injuries and Illnesses in cooperation with participating state agencies. Here we consider the distribution of nonfatal injuries resulting in at least one day lost from work. Table I provides the raw data used for our study [2] for 28 different labor categories. The first row is the combined set of data for all private industry in the U.S.

The power-law distribution of Eq. (1) is independently fit to each row of data in Table I to determine the optimized values of $a$ and $k$ using a least squares method. For the columns in Table I where the reported data are binned across multiple days, we consider an average number of accidents at the midpoint of the reported range of days. The final column of data for 31 or more lost days of work is not used for curve fitting, since the upper bound of the number of days is unknown and a midpoint cannot be determined. The quality of the power-law fits is quantified by the standard coefficient of determination, $R^2$. The optimized parameters, $a$ and $k$, and the $R^2$ values for each labor category are provided in Table II.

## III. Results

The power-law distribution provides an excellent fit to all of the workplace accident data across all labor categories, as the $R^2$ values are above 0.9 in all cases and above 0.95 for most categories. Since the $R^2$ values represent the proportion of the variance in the workplace accident rates that are predictable from the power-law distribution, more than 90% of the variation in accident rates with severity can be explained by a power-law model. Figure 1 plots the power-law distribution for the total number of workplace accidents across all of private industry. The $R^2$ value for the power-law fit is 0.977, indicating that it is an excellent description of the workplace accident distribution.

Figure 2 shows power-law distributions for two of the individual labor categories from Table I: "Servicing Providing" and "Management of Companies and Enterprises." Both sectors exhibit power-law scaling with $R^2 > 0.975$. The slopes are nearly identical, with $k \approx 1.02$ in both cases, indicating that the average number of days lost from a particular workplace accident is roughly the same in both labor categories. The only difference in the two distributions is a vertical shift in Fig. 2 owing to the much larger workforce population in the "Service Providing" sector. The fact that the workplace accident data in both of these independent labor sectors—and in every category under study—are so well described by power-law distributions is strong evidence for the self-organized nature of workplace accident statistics. Indeed, the high $R^2$ values for the power-law fits in every category in Table II indicates that the power-law distribution emerges naturally for any category of labor.

While most of the labor categories exhibit power-law scaling of workplace accidents with $k$ values near 1.0, a few categories show moderate departure either above or below this exponent value. Two of these are plotted in Fig. 3, which shows the distribution of workplace accidents for the "Arts, Entertainment, and Recreation" category, exhibiting a higher exponent of $k = 1.147$, and for the "Mining" category, which has a lower exponent of $k = 0.761$. A smaller exponent indicates a greater average number of days lost in a workplace accident; hence, individual mining accidents are, on average, more severe than accidents that occur in the arts, entertainment, and recreation industry. However, it is important to note that even with these differences in slope, a power-law distribution with high $R^2$ values emerges for both of these completely independent and unrelated job categories.

## IV. Discussion

The power law scaling of workplace accident frequency with severity across all job categories considered by the U.S. Bureau of Labor Statistics is a strong indication that workplace accidents are governed by self-organized criticality. The practical implication of this result is that nearly all workplace accidents have a common underlying cause, i.e., the same conditions, attitudes, behaviors, etc., that lead to common (minor) workplace injuries also yield the less frequent major injuries. Thus, in order to suppress the occurrence of major accidents, we must focus on reducing the entire distribution of accident rates. By reducing the rate of minor accidents, the entire distribution will be suppressed, i.e., the rate of major accidents and fatalities will decrease.

If all workplace accidents have a common underlying origin, then the only way to effectively reduce accident rates is to identify and address this cause. Heinrich's studies have shown that workplace accidents can be traced back to unsafe acts related to job hazards or worker behaviors [12]. The existence of such unchecked hazards and unsafe actions is a direct result of the culture of the organization. Does the organization value safety, and does it translate that value into action? To reduce accident rates overall, organizations must strive to create a workplace culture where safety is one of its core values among all levels of management, with top management setting expectations and leading by example, and among workers in all roles in the organization.

In creating this culture where safety is a core value, particular emphasis should be placed on identifying and addressing the "minor" issues that are systemic of a potentially dangerous workplace. These minor issues, which can be as seemingly benign as issues related to housekeeping and organization, occur the most frequently in the workplace and are therefore the best candidates for addressing directly through proper awareness and training. By creating a culture that recognizes and rectifies even small safety-related concern, the accident rates should decrease across the full range of severities. Creating such a culture is a journey, not a destination, in that continuous improvement is necessary, with a belief and attitude toward safety that stretches beyond simply being compliant. This approach is also consistent with the desire to move from "lagging" indicators such as accident rate statistics toward "leading" indicators such as the predictive assessment of the safety climate of an organization [13].

Our results are consistent with the Heinrich's pioneering research into workplace accident statistics [12]. Heinrich proposed an "accident triangle" describing how the number of severe accidents occurs much less frequently compared to that of minor incidents. Heinrich proposed that nearly all accidents have a common underlying cause related to unsafe actions and hazards in the workplace. Heinrich's work has had an enormous influence on thinking around workplace safety for many decades, but recently it has been criticized by some as being "unscientific" [14]. For example, Anderson and Denkl have argued that Heinrich's Law only applies to aggregate data and not to individual activities and that separate efforts should be specifically devoted to high-risk activities [15]. Our results herein give scientific support to Heinrich's postulate of a common origin for workplace accidents since the accident statistics in all occupation categories—including higher-risk occupation such as mining—demonstrate a power law distribution indicative of self-organized criticality.

## V. Conclusions

The statistics of rare events is a major concern for a variety of applications in physics, biology, geology, and across a wide range of social sciences. While initially developed to describe the distribution of landslides occurring in sandpiles, self-organized criticality has found far-reaching applications throughout the natural and social sciences. A key feature of self-organized criticality is the emergence of power-law scaling for the frequency of events as a function of size or severity, independent of the specific parameters of the experiment. In this work we have considered the statistics of workplace accident frequency as a function of the time lost from work for 28 different labor categories, as tabulated by the U.S. Bureau of Labor Statistics. In every category, the distributions are well described by a power-law model; the emergence of power-law scaling independent of labor category is a strong indication that workplace accidents are the result of self-organized criticality. This points to a common underlying cause for nearly all workplace accidents, irrespective of their severity. This result is analogous to that obtained, e.g., for the distribution of landslide sizes in sandpiles or the frequency of earthquakes as a function of seismic magnitude. Just as self-organized criticality explains that all landslides and earthquakes have a common underlying cause, independent of their magnitude, we can also arrive at the practical conclusion that workplace accidents are caused by common underlying factors.

Thus, in order to reduce the rate of major accidents and fatalities, a workplace culture needs to be advanced that addresses *all* accidents, not just the most severe. By focusing on creating an attitude and culture of safety and addressing common minor problems such as housekeeping and cleanliness issues, consistent proper use of personal protective equipment, etc., the rate of accidents as a whole will be brought down. We hope that the insights provided by this physics-bases analysis and interpretation will provide an explanation for *why* focusing on creating a safety-oriented culture is the most effective strategy for reducing accident rates of all levels of severity, in agreement with the arguments made by Heinrich and other researchers in the safety science community [16]-[17].

**Acknowledgements:** We are grateful for valuable discussions with E. Kupp, S. Henninger, and D.C. Allan.

**Table I.** Number of workplace accidents leading to different number of days away from work [2].

| | Total Cases | Days Away from Work Cases Involving | | | | | | |
| | | 1 | 2 | 3 to 5 | 6 to 10 | 11 to 20 | 21 to 30 | 31 or More |
|---|---|---|---|---|---|---|---|---|
| **Private Industry** | 902,160 | 127,610 | 100,510 | 156,430 | 104,900 | 96,590 | 57,800 | 258,330 |
| **Goods Producing** | 226,320 | 30,660 | 23,050 | 36,930 | 25,030 | 25,450 | 15,060 | 70,140 |
| **Natural Resources and Mining** | 23,830 | 2,900 | 2,810 | 4,230 | 3,390 | 2,370 | 1,520 | 6,610 |
| **Agriculture, Forestry, Fishing, and Hunting** | 18,660 | 2,490 | 2,540 | 3,750 | 2,700 | 1,890 | 1,170 | 4,130 |
| **Mining** | 5,160 | 410 | 270 | 470 | 690 | 480 | 350 | 2,490 |
| **Construction** | 79,890 | 9,780 | 7,520 | 12,470 | 7,920 | 9,060 | 5,410 | 27,720 |
| **Manufacturing** | 122,610 | 17,980 | 12,720 | 20,230 | 13,720 | 14,020 | 8,130 | 35,800 |
| **Service Providing** | 675,840 | 96,950 | 77,460 | 119,500 | 79,870 | 71,130 | 42,740 | 188,190 |
| **Trade, Transportation, and Utilities** | 279,150 | 35,360 | 26,830 | 44,560 | 31,780 | 31,110 | 18,810 | 90,700 |
| **Utilities** | 4,060 | 280 | 240 | 610 | 440 | 600 | 310 | 1,580 |
| **Wholesale Trade** | 60,340 | 8,790 | 6,570 | 9,960 | 6,570 | 6,450 | 3,720 | 18,290 |
| **Retail Trade** | 123,770 | 19,550 | 14,150 | 22,470 | 14,050 | 12,900 | 8,150 | 32,510 |
| **Transportation and Warehousing** | 90,990 | 6,750 | 5,870 | 11,520 | 10,720 | 11,160 | 6,640 | 38,330 |
| **Information** | 14,050 | 1,300 | 1,300 | 1,900 | 1,490 | 1,440 | 770 | 5,860 |
| **Financial Activities** | 30,110 | 4,090 | 3,920 | 5,330 | 3,600 | 2,020 | 1,680 | 9,470 |
| **Finance and Insurance** | 9,480 | 1,600 | 1,120 | 1,140 | 740 | 730 | 940 | 3,200 |
| **Real Estate and Rental and Leasing** | 20,630 | 2,490 | 2,800 | 4,190 | 2,860 | 1,290 | 740 | 6,270 |
| **Professional and Business Services** | 67,320 | 10,600 | 8,160 | 12,060 | 7,190 | 6,310 | 4,430 | 18,550 |
| **Professional and Technical Services** | 18,030 | 4,340 | 2,350 | 3,170 | 2,070 | 940 | 1,310 | 3,860 |
| **Management of Companies and Enterprises** | 5,700 | 1,000 | 510 | 950 | 720 | 820 | 310 | 1390 |
| **Administrative and Waste Services** | 43,590 | 5,260 | 5,310 | 7,940 | 4,400 | 4,550 | 2,820 | 13,310 |
| **Educational and Health Services** | 168,940 | 26,740 | 21,660 | 32,500 | 22,350 | 17,940 | 9,470 | 38,270 |
| **Educational Services** | 10,530 | 1,820 | 1,650 | 1,880 | 1,240 | 1,270 | 410 | 2,270 |
| **Health Care and Social Assistance** | 158,410 | 24,930 | 20,010 | 30,620 | 21,110 | 16,670 | 9,060 | 36,010 |
| **Leisure and Hospitality** | 92,670 | 15,170 | 12,340 | 18,610 | 10,850 | 9,520 | 5,960 | 20,220 |
| **Arts, Entertainment, and Recreation** | 14,110 | 2,630 | 1,790 | 2,530 | 1,810 | 1,510 | 680 | 3,180 |
| **Accommodation and Food Services** | 78,560 | 12,540 | 10,550 | 16,080 | 9,040 | 8,010 | 5,290 | 17,040 |
| **Other Services, Except Public Administration** | 23,600 | 3,690 | 3,250 | 4,530 | 2,610 | 2,800 | 1,600 | 5,120 |

**Table II.** Parameters for optimized power-law fits using Eq. (1) and the data provided in Table I. $R^2$ is the coefficient of determination and represents the proportion of the variance in the workplace accident rate that is predictable from the power-law distribution.

| | $a$ | $k$ | $R^2$ |
|---|---|---|---|
| **Private Industry** | 168,352 | 1.014 | 0.977 |
| **Goods Producing** | 38,732 | 0.978 | 0.982 |
| **Natural Resources and Mining** | 4,306 | 0.989 | 0.951 |
| **Agriculture, Forestry, Fishing, and Hunting** | 3,871 | 1.034 | 0.946 |
| **Mining** | 456 | 0.761 | 0.957 |
| **Construction** | 12,268 | 0.944 | 0.980 |
| **Manufacturing** | 22,130 | 0.996 | 0.986 |
| **Service Providing** | 129,696 | 1.026 | 0.976 |
| **Trade, Transportation, and Utilities** | 44,838 | 0.952 | 0.981 |
| **Utilities** | 366 | 0.688 | 0.930 |
| **Wholesale Trade** | 11,229 | 1.027 | 0.983 |
| **Retail Trade** | 24,912 | 1.043 | 0.982 |
| **Transportation and Warehousing** | 8,713 | 0.740 | 0.960 |
| **Information** | 1,871 | 0.928 | 0.957 |
| **Financial Activities** | 6,220 | 1.117 | 0.948 |
| **Finance and Insurance** | 1,675 | 1.023 | 0.942 |
| **Real Estate and Rental and Leasing** | 4,614 | 1.194 | 0.914 |
| **Professional and Business Services** | 13,827 | 1.067 | 0.977 |
| **Professional and Technical Services** | 4,908 | 1.228 | 0.955 |
| **Management of Companies and Enterprises** | 1,105 | 1.020 | 0.980 |
| **Administrative and Waste Services** | 7,727 | 1.001 | 0.954 |
| **Educational and Health Services** | 37,537 | 1.083 | 0.970 |
| **Educational Services** | 2,732 | 1.189 | 0.959 |
| **Health Care and Social Assistance** | 34,841 | 1.077 | 0.970 |
| **Leisure and Hospitality** | 20,819 | 1.080 | 0.972 |
| **Arts, Entertainment, and Recreation** | 3,437 | 1.147 | 0.982 |
| **Accommodation and Food Services** | 17,386 | 1.069 | 0.969 |
| **Other Services, Except Public Administration** | 5,063 | 1.045 | 0.971 |

**FIGURES**

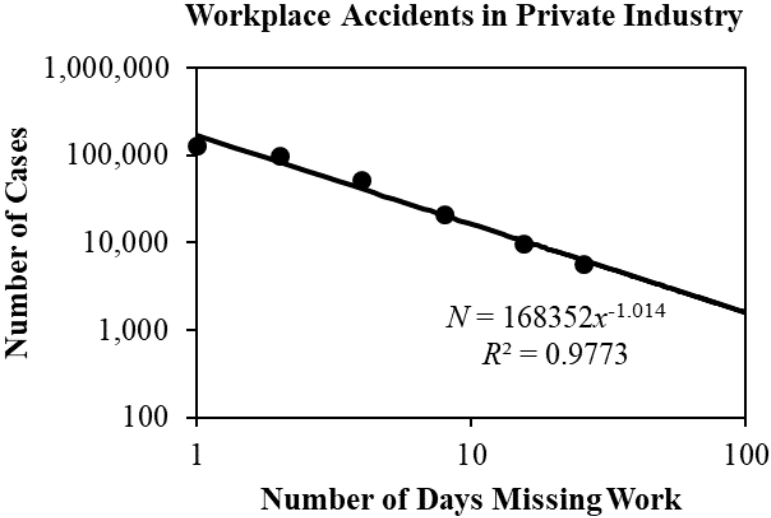

**Figure 1.** Power-law distribution of the number of workplace accidents in all private industry as a function of the number of days missed from work.

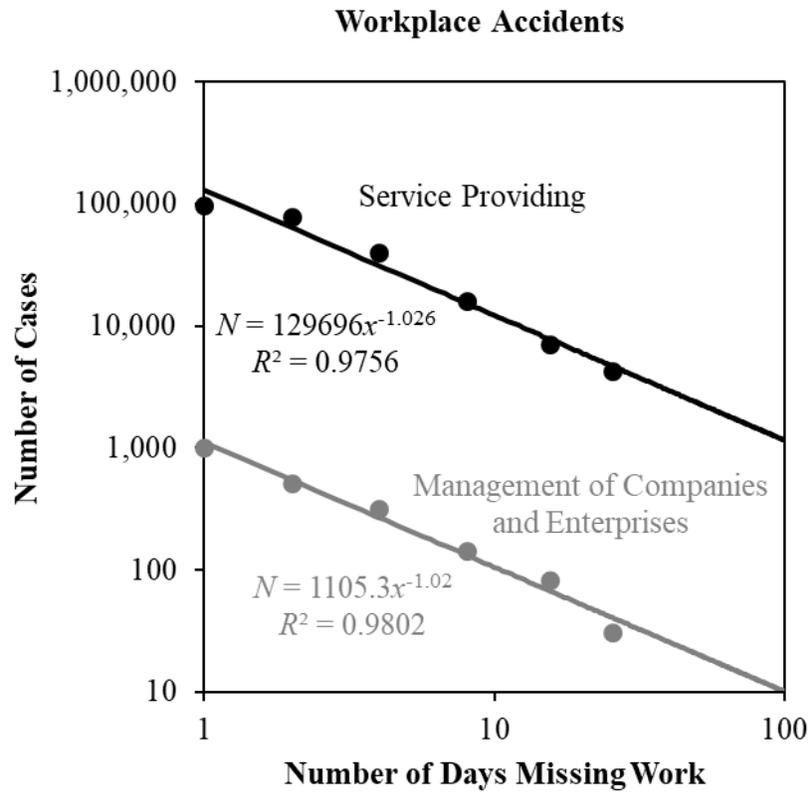

**Figure 2.** The distributions of workplace accidents in the "Servicing Providing" and "Management of Companies and Enterprises" sectors exhibit power-law scaling with nearly identical exponents.

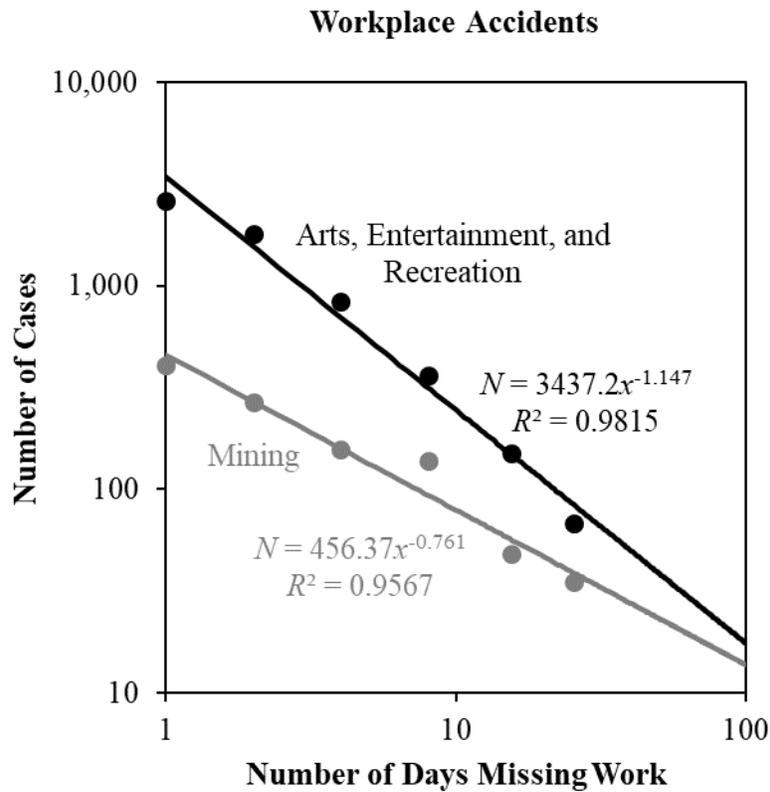

**Figure 3.** Power-law distribution of the number of workplace accidents in the "Arts, Entertainment, and Recreation" category, which exhibits a large exponent, compared to number of workplace accidents in the "Mining" category, which exhibits a smaller exponent. A smaller exponent indicates a greater average number of days lost in mining accidents compared to accidents in the arts, entertainment, and recreation industry.